\documentclass[12pt]{article}

\usepackage{graphicx}
\usepackage{epstopdf}
\usepackage[body={15.5cm, 21cm},right=3cm]{geometry}
\usepackage{amssymb}
\usepackage{amsmath}

\newcommand\eea{\end{eqnarray}}
\newcommand\bea{\begin{eqnarray}}

\def\beq{\begin{equation}}
\def\eeq{\end{equation}}

\newcommand{\be}{\begin{equation}}
\newcommand{\ee}{\end{equation}}
\newcommand{\ba}{\begin{align}}
\newcommand{\ea}{\end{align}}
\newcommand{\bg}{\begin{gather}}
\newcommand{\eg}{\end{gather}}
\newcommand{\bseq}{\begin{subequations}}
\newcommand{\eseq}{\end{subequations}}

\begin{document}

\begin{flushright}
NSF-KITP-12-027,  ~~~~ SLAC-PUB-14883, ~~~~ SU-ITP-12/08 
\end{flushright}

\vspace{3mm}
\vspace{0.5cm}
\begin{center}

\def\thefootnote{\fnsymbol{footnote}}

{\Large \bf Large-density field theory, viscosity, and ``$2k_F$" singularities from string duals}
\\[0.5cm]
{\large  Joseph Polchinski$^1$ and Eva Silverstein$^{1,2}$}
\\[0.5cm]

{\small \textit{$^{1}$
KITP\\
University of California\\
Santa Barbara, CA,  USA}}

{\small \textit{$^{2}$
Department of Physics and SLAC \\
Stanford University, Stanford, CA 94305, USA}}

\vspace{.2cm}

%{\small \textit{$^{2}$ Institute for Advanced Study\\Einstein Drive\\ Princeton, NJ 08540}}

\end{center}

\vspace{.8cm}

\hrule \vspace{0.3cm}
{\small  \noindent \textbf{Abstract} \\[0.3cm]
\noindent
We analyze systems where an effective large-N expansion arises naturally in gauge theories without a large number of colors: a sufficiently large charge density alone can produce a perturbative string ('tHooft) expansion.  One example is simply the well-known NS5/F1 system dual to $AdS_3\times T^4\times S^3$, here viewed as a 5+1 dimensional theory at finite density.  This model is completely stable, and we find that the existing string-theoretic solution of this model yields two interesting results.  First, it indicates that the shear viscosity is not corrected by $\alpha'$ effects in this system.  For flow perpendicular to the F1 strings the viscosity to entropy ratio take the usual value $1/4\pi$, but for flow parallel to the F1's it vanishes as $T^2$ at low temperature.  Secondly, it encodes singularities in correlation functions coming from low-frequency modes at a finite value of the momentum along the $T^4$ directions.    This may provide a strong coupling analogue of finite density condensed matter systems for which fermionic constituents of larger operators contribute so-called ``$2k_F$" singularities.
In the NS5/F1 example, stretched strings on the gravity side play the role of these composite operators.  We explore the analogue for our system of the Luttinger relation between charge density and the volume bounded by these singular surfaces.  This model provides a clean example where the string-theoretic UV completion of the gravity dual to a finite density field theory plays a significant and calculable role.

\vspace{0.5cm}  \hrule
\def\thefootnote{\arabic{footnote}}
\setcounter{footnote}{0}

%\def\be{\begin{eqnarray}}
%\def\ee{\end{eqnarray}}

%\date{\today}

%\maketitle

\vspace{.8cm}

%\tableofcontents

\vspace{.8cm}
\baselineskip = 16pt

\section{Introduction}

The duality between Yang-Mills theory with a large number $N_c\gg 1$ of colors and perturbative string theory has much improved our understanding of quantum gravity and of quantum field theory.  A basic question which remains is to characterize which strongly coupled systems should admit a gravity dual, or at least a perturbative string theory dual.  In this note we consider examples where the latter arises not because of large $N_c$, but because of large density.

At large density $n$, a system of particles with a weak coupling constant may interact strongly simply because the particles scatter at a large rate
%\beq\label{sigmanv}
$\Gamma \sim \sigma n v$
%\eeq
where $v$ is the typical particle velocity.
In low temperature quantum condensed matter systems, this scattering picture is not appropriate but the same question arises:  could strong coupling arise simply because of finite density effects in an otherwise weakly interacting system?  If so, under what conditions is this strong coupling physics controlled by an effective 'tHooft expansion?

In simple holographic models, we will see that the existence of a perturbative string description does not depend on the underlying theory being a large-$N_c$ gauge theory.  Using the original strategy \cite{Maldacena:1997re}, one can obtain a non-gravitational low-energy theory by taking a near-horizon limit of a higher-dimensional brane system with strong gravitational redshift.  Generically, there are multiple sectors of branes in such a construction, each extended over some directions and distributed along others.  Some subset of these branes contribute a density of lower-dimensional charges, as opposed to contributing Yang-Mills ``colors" (adjoint gauge fields and other matter propagating in the in the $d$ dimensions of the field theory) or ``flavors" (such as fundamental matter representations).
% We will refer to these various sectors as color branes, flavor branes and density branes.
Moreover, one may isolate the very low energy theory, remaining agnostic about its UV completion.
%couple it to $d$-dimensional fields in a ``semiholographic" construction, which still retains much of the power of the %correspondence \cite{Semihol}.
As we will see shortly, it is straightforward to determine under what conditions the density branes contribute dominantly to the gravitational redshift  and produce a controlled supergravity or classical string theory solution.  One example which we will focus on is in fact a classic example of AdS/CFT, built from a combination of 1-branes and 5-branes, simply treating it as a $5+1$ dimensional theory at finite density.

This question may be of particular interest in attempting to apply the holographic correspondence to condensed matter problems, where large-$N_c$ Yang-Mills theory does not arise naturally, but finite density is crucial.\footnote{Indeed, from a different (reasonably realistic) starting point, S.-S. Lee suggested a specific way in which finite-density fermion physics may lead to a 'tHooft expansion in a theory with $N_c=1$ \cite{Lee:2010fy}.  Although that example turned out to hold further subtleties that may obstruct a simple 'tHooft expansion \cite{MetSach}, the general idea works in other examples as we will see.}
One basic question that has arisen in condensed matter theory has to do with so-called ``$2k_F$" singularities.  In weakly interacting fermi liquids, there are low-frequency, finite-momentum excitations of particles and holes.  In a free theory, this spectrum truncates at twice the fermi momentum, leading to singularities in the density-density two point function.  Various arguments and calculations indicate that this behavior persists beyond the free theory \cite{Luttinger},  extending to situations where the underlying fermions are constituents of larger gauge-invariant operators.
At strong coupling, the traditional field theoretic calculations are more difficult to control, and it is interesting to ask whether singularities from soft modes at nonzero momentum occur for holographic field theories.

Progress on strongly coupled Fermion physics was made by introducing probe fermions in a much larger finite density field theory \cite{MIT}, yielding theories with fermi surfaces and non-fermi liquid transport in the probe sector (before including various instabilities).  But in these cases \cite{MIT}\ and others \cite{HPST}\ with a simple mechanism for non-Fermi-liquid transport, the leading supergravity contribution to the density density correlator computed on the gravity side (at large radius and weak coupling) does not exhibit singularities in momentum (see \cite{AdStwo}\cite{Puletti:2011pr}\ and \cite{KulaxiziParnachev}\ respectively).

As we will discuss further below, exhibiting any singularities in momentum that do persist at strong coupling might generically require string theoretic or quantum corrections, since it may require exposing particle/hole pairs.  In particular, there have been suggestions that in some of these examples string-theoretic corrections could become important and be sufficient to introduce such effects, given a stable and controllable solution.\footnote{One example may come from the infrared region of DBI flavor branes, where ultimately string-theoretic effects become important \cite{unpubstr}\cite{HPST}.}

\subsection{Lord of the Strings fan physics}

In exploring large-density field theories, we find that a classic example of AdS/CFT -- the subject of the epic trilogy \cite{AdSthree}\ and other important works such as \cite{Hobbit}\cite{Silmarillion}\cite{AdSthreesusy} -- realizes this idea in a precise way.
Namely, the infrared limit of the system of $N_5$ Neveu-Schwarz 5-branes and $N_1$ fundamental strings (extended along one of the five spatial directions, and smeared in the other four) is dual to $AdS_3\times R^4 \times S^3$.  This has mostly been viewed as an example of $AdS_3/CFT_2$ duality.  It is, however, equally valid to describe it as a 5+1 dimensional non-gravitational theory at a finite density of 1-dimensional extended objects (related to instanton strings in an appropriate limit).  We will focus on the infrared region of this system.  The case $N_1=0$ was analyzed in a similar spirit in the work \cite{Andreis}.  
%coupling it semiholographically to the gauge field under which the strings are charged (in field theory language, weakly %gauging that symmetry).
Using the beautiful solution of this theory developed in \cite{Silmarillion}\cite{Hobbit}\cite{AdSthree},  we can read off very precisely the stringy corrections to transport, and to the more detailed structure of correlation functions.  We find three main results: 

 (1) For flow perpendicular to the strings, the shear viscosity (a zero momentum quantity) is equal to the entropy density times $1/4\pi$, uncorrected in classical string theory (as found earlier for $N_1=0$ \cite{Andreis}).
 
  (2)  For flow parallel to the strings, the shear viscosity to entropy ratio {\it vanishes} as $T^2$ at low temperature.  Thus, given the anisotropic nature of this system,  it approaches a perfect liquid crystal.
  
(3) There are singularities in momentum which go down to arbitrarily small frequency, realizing an analogue of ``$2k_F"$ singularities in this strongly coupled system.

Previous studies of the shear viscosity tensor in anisotropic holographic systems include \cite{Landsteiner:2007bd}\cite{Natsuume:2010ky}\cite{Erdmenger:2010xm}\cite{Basu:2011tt}\cite{Erdmenger:2011tj}\cite{Rebhan:2011vd}.
Refs.~\cite{Erdmenger:2010xm}\cite{Basu:2011tt}\cite{Erdmenger:2011tj}\cite{Rebhan:2011vd} found nonuniversal values for certain components of this tensor.  In particular, Ref.~\cite{Rebhan:2011vd}  found a value less than the usual universal bound, in the anisotropic plasma based on dissoved branes introduced in \cite{Mateos:2011ix}. 

The supersymmetry of this example --- while excessive in comparison to real world systems --- is very useful in establishing the stability of the background.
A complication in finite density holography is that it has been difficult to establish UV complete solutions extending arbitrarily far in the infrared including backreaction; various interesting instabilities and singularities arise which remain to be fully understood.  Recent progress was made in \cite{Almuhairi:2011ws}\ where a supersymmetric magnetic brane system leads to $AdS_2\times R^2$.\footnote{Some arguments \cite{Jensen:2011su}\ suggest that any $AdS_2$ geometry should ultimately be unstable in the deep infrared.}

\section{Large density expansions:  general considerations}
\setcounter{equation}{0}

To begin, let us check the conditions under which large density can drive a controlled gravity or string theory description of a field theory.  To this end, we will analyze the contributions of various brane sources to the curvature radius and string coupling.  Let us consider systems with approximate translation invariance in some number $p$ of the spatial directions, with a finite density of $d_\rho$-dimensional objects distributed nearly uniformly along $p$ directions.  We will view this as a candidate $p+d_\rho+1$ dimensional field theory at finite density.  For simplicity we will start by exploring a density of pointlike charges ($d_\rho=0$), considering D0-branes in type IIA string theory, but our main example will involve a density of fundamental strings ($d_\rho=1$).

The simplest way to introduce a small-$N$ gauge theory sector is semiholographically \cite{Semihol}, weakly gauging the flavor symmetry under which the density branes are charged.  More generally, we can include a combination of $N_c$ color $(p+d_\rho)$-branes along with the density branes (leaving out flavor branes here just for simplicity), and ask under what circumstances a controlled dual may be obtained with $N_c$ not much greater than 1.

The effective action in superstring theory has the schematic form (in the string frame)
\beq\label{action}
{\cal S}=\int \frac{d^{10} x}{{\alpha'}^4} \left(\frac{1}{g_s^2}{\cal R}+\frac{1}{g_s^2}|H_{(3)}|^2+\sum_{\tilde p}|F_{({\tilde p})}|^2+\dots \right) \,.
\eeq
Here $g_s\sim e^{\Phi}$ is the string coupling, related to the dynamical dilaton field $\Phi$.  ${\cal R}$ is the curvature scalar, $H_{(3)}=dB_{(2)}$ is the field strength corresponding to a two-form Neveu-Schwarz (NS) gauge potential $B_{(2)}$, and $F_{(\tilde p)}=dC_{({\tilde p}-1)}$ denote Ramond-Ramond (RR) field strengths.  The values of $\tilde p$ which appear depends on the particular version of string theory under study (odd for type IIA, even for type IIB).  D-branes are charged under RR gauge fields, while fundamental strings and NS 5-branes are electrically and magnetically charged under the NS gauge field.  In the gravity solutions for the branes, the corresponding fluxes are therefore turned on, a feature which will be important momentarily.

It is straightforward to use the scalings of the various source terms in the action to determine under what circumstances we can obtain a controlled holographic large-density theory in $p+1$ dimensions without a large color sector.  For this purpose we may simply work in in the string frame, and we will not consider situations with any finely tuned cancellations among different sources.  We have a curvature term
\beq\label{curvature}
\frac{1}{g_s^2}{\cal R}\sim \frac{1}{g_s^2 R^2} \,,
\eeq
where $R$ is the curvature radius of the solution in units of the string tension.
A density of D0-branes smeared along $p$ directions leads to a term of order
\beq\label{Dzero}
|F_{(8)}|^2\sim \frac{\rho_{0}^2}{R^{2(8-p)}}
\eeq
in the gravity solution.  Here $\rho_0=N_0/L_p^p$ (with $L_p$ the size in string units of the $p$ dimensions along which the D0-branes are smeared).   Similarly, a density of fundamental strings smeared along $p$ directions gives a term
\beq\label{Str}
\frac{1}{g_s^2}|H_{(7)}|^2\sim \frac{g_s^2\rho_{F1}^2}{R^{2(7-p)}}  \,,
\eeq
where $\rho_{F1}=N_{F1}/L_p^p$ and $H_{(7)}$ is the field strength magnetically dual to $H_{(3)}$.
Unsmeared color D$p'$-branes contribute a term
\beq\label{Dp}
|F_{(8-p')}|^2\sim \frac{N_{Dp'}^2}{R^{2(8-p')}} \,,
\eeq
while NS5-branes contribute a term scaling like
\beq\label{NS}
\frac{1}{g_s^2}|H_{(3)}|^2\sim\frac{N_{NS5}^2}{g_s^2 R^6} \,.
\eeq
A perturbative string description requires $g_s\ll 1$, and a general relativistic description requires further that the curvature radius in string units be large, $R\gg 1$.  This happens at all scales in solutions with an $AdS$ factor, but it is also useful if it occurs over a large but finite range of scales (corresponding to a large radial distance in the corresponding gravity solutions) \cite{Itzhaki:1998dd}.

Consider first a smeared density of D0-branes \cite{Smeared}.  Matching the terms (\ref{Dzero}) and (\ref{curvature}), we find that any solution in a large-radius regime will satisfy
\beq\label{Dzerolarge}
R^{7-p}\sim g_s\rho_0 \,.
\eeq
This kind of relationship is familiar in AdS/CFT (even for non-conformal branes), but with the number of colors $N_c$ playing the role that is played here by $\rho_0$, the proper density of D0-branes on the gravity side in units of the string tension $1/\alpha'$.
%That is, $\rho_0=N_0\alpha'/L^p$, where $L$ is the proper size of the dimensions along %which the D0-branes are smeared (taken compact for the present purposes).
This dimensionless combination can also be described on the field theory side, by considering the dual description of  a string state to incorporate the $\alpha'$ scale that appears here.\footnote{It seems interesting that in the dense D0-brane plasma, strings correspond to current loops.  (This is the T-dual of electric flux, which in AdS/CFT is the field theory dual of strings for the case of Dp color branes \cite{Balasubramanian:1998de}.)}
%In the dual field theory, we may measure distances in the $p$ spatial directions according to whatever convention we wish, %and physical quantities such as $\rho_0$ are independent of this choice.
%One way to organize this is to carry over the gravity side quantities fairly directly.
%For simplicity, we may use units in which the proper distances on the gravity side are equal to those in the field %theory at the radial position corresponding to the scale at which we probe the system.   In proper units, a string %propagating on the gravity side have a size of order $\sqrt{\alpha'}$. The field theory operators which create strings %moving through this position have support (energy density) concentrated over a distance of order $\sqrt{\alpha'}$ in %the $p$ spatial directions.  The dimensionless product $\rho_0$ of the density and a volume of this size is %independent of how we measure spatial distances in the field theory, and is meaningful.
In our main example below, the NS5/F1 system, we will find a simple field theoretic way to characterize this by relating the density to the position of a certain singularity in momentum space
that can be read off from correlation functions.

The relation (\ref{Dzerolarge}) suggests that density itself can form the basis of a controlled expansion, a statement borne out by the supergravity solution.  In string frame, the near horizon solution is
\bea\label{Dzeromet}
ds^2 &=& -H^{-1/2} f dt^2 + H^{1/2} (f^{-1}dr^2+r^2 d\Omega^2 + d\vec x^2) \nonumber\\
e^{2\phi} &=& H^{3/2}g_0^2 \nonumber\\
A &=& \coth\beta (1-H^{-1})dt
\eea
with
\beq
H(r) = \frac{r_0^{7-p}\sinh^2\beta}{r^{7-p}} ~~~~~~ f=1-\frac{r_0^{7-p}}{r^{7-p}} \,.
\eeq
Here the constants $\beta$ and $r_0$ are related to the charge density and temperature of the dual theory \cite{Smeared}.
The Hamiltonian describing this plasma at zero temperature is the maximally supersymmetric matrix quantum mechanics, and the existence of the supergravity solution suggests that there is a threshold bound state with the symmetries of the above solution, analogously to the supergravity solution for localized D0-branes \cite{Itzhaki:1998dd}, which form a threshold bound state \cite{BoundState}.

One interesting example is the $p=3$ case, for which the metric in string frame is
\beq\label{Dzerothree}
ds^2 = -\frac{r^2}{R^2} f dt^2 + \frac{R^2}{r^2}d\vec x^2 + \frac{R^2}{fr^2} dr^2 + R^2 d\Omega^2
\eeq
where $R^2 = r_0^2 |\sinh\beta| $.
This metric is very similar to the black D3-brane metric, with the distinction that the three spatial directions coordinatized by $\vec x$ here grow toward the IR (small $|g_{tt}|$) rather than shrinking.  (Also, the dilaton runs in the smeared D0 solution, unlike in the D3-brane solutions.)  At nonzero temperature, the system may have instabilities toward clumping of the matrix eigenvalues as indicated by the near-horizon version of the Gegory-Laflamme instability \cite{AharonyWiseman}, something that would be interesting to see directly in the dual field theory.

%It would be interesting to analyze this theory further, for example weakly gauging a flavor symmetry and computing the plasma-screened gauge field propagator.

So far we have $N_c=0$ here, but
we may obtain an effective small-$N_c$ gauge theory by weakly gauging a flavor symmetry in the D0-brane system, that is semi-holographically coupling  one or more of its flavor currents to gauge fields and computing the plasma effects on the gauge field propagators
(at least at zero temperature).  Doing this for a RR gauge field (which does not couple to the dilaton classically)  we obtain a system which is similar to the study in \cite{Plasma}\ of the plasma physics of the $N=4$ SYM theory; the main difference arises for nonzero momentum modes along the $\vec x$ directions.  In the case of the D0-brane plasma, the momentum modes are unsuppressed as we go toward the IR.

If we add $N_p$ color Dp-branes to the D0-brane density smeared along $p$ directions, the two compete with each other and with curvature if $N_p\sim \rho_0\sim R^{7-p}/g_s$.  This requires a large-color theory for holographic control.
A similar example is the D1/D5 theory, for which $R^2\sim g_s N_{D5}$ and $N_{D5}\sim \rho_{D1}$.  In these examples, although large $N_c$ is required to obtain a large radius, the holographic control of the infrared behavior is driven by the large density.  In the absence of the density of D1-branes, the field theory becomes weakly coupled in the infrared, and the gravity dual correspondingly becomes highly curved.  The density of D1-branes prevents this breakdown of the geometry.

If rather than allowing the two terms to compete, we keep the D0 density dominant (considering a regime with $\rho_0\gg N_p$), then we go beyond the usual near-horizon limit of the Dp-brane color sector, and the color branes are probes of the geometry dual to the D0-brane density.  In this regime, in other words, the D0-brane density is still the dominant effect causing the brane system to decouple from gravity.

If we add $N_{5}$ NS5-branes to a density of D0-branes smeared over $p=5$ directions, we obtain a solution with (over a range of scales) $R^2\sim N_{5}$ and $g_s\sim N_{5}/\rho_0$.  Thus we obtain a weakly coupled string theory even if we consider a system with a small number of NS5-branes.  A solvable analogue of this is what we will consider in detail in the remainder:  the combination of $N_5$ NS5-branes and $N_1$ F1-strings.  This yields
\beq\label{NSFonescalings}
\hat R^2\sim N_5\alpha'\,,\quad g_s^2\sim \frac{N_5}{\hat \rho_1 {\alpha'}^2}
\eeq
where in this expression we make the dimensions manifest; $\hat R=R\sqrt{\alpha'}$ and $\hat\rho_1=\rho_1/{\alpha'}^2$ are the curvature radius and fundamental string density.
Again in this example a weakly coupled string theory limit is obtained at large density.  As described above in the case of smeared D0-branes below (\ref{Dzerolarge}), the criterion $\hat\rho_1{\alpha'}^2\gg 1$ can be described also in field theory terms.  We will return to this below, finding a simpler way to describe this dimensionless quantity in terms of singularities of correlation functions in momentum space.

\section{NS5-F1, viscosity and singularities in momentum space}
\setcounter{equation}{0}

This last example is of course one of the original examples~\cite{Maldacena:1997re} of AdS/CFT duality.  Our point here is that it is useful to view the low energy NS5/F1 system as a 5+1 dimensional non-gravitational theory at finite density.   In particular, we will see that the solution \cite{AdSthree}\ encodes interesting physics involving singularities in momentum space.

Let us first briefly review the system of $N_5$ NS5-branes on a $T^4$ and $N_1$ fundamental strings smeared over the $T^4$.  At high energies, this is an exotic non-gravitational 5+1-dimensional theory known as ``little string theory" \cite{LST}.  This is an interesting UV completion of the low energy theory that we will analyze below, but it is more exotic than we need for our application and we will not make use of it.  For sufficiently large $N_5$, there is a range of scales where the theory is a 5+1 dimensional Yang-Mills theory.

In the deep infrared, the system reduces to string theory on the target spacetime $AdS_3\times T^4\times S^3$, with $N_5$ units of magnetic 3-form NS-NS field strength $H_{ijk}$ on the $S^3$ and $N_5$ units of electric flux $H_{012}$ on $AdS_3$, with radius $R_{AdS}\sim\sqrt{N_5}$ and with string coupling $g_s\sim (N_5/\rho_1)^{1/2}$ where $\rho_1=N_1/V_{T^4}{\alpha'}^2$ is the F1 density on the $T^4$.  This is an exactly solvable theory at the classical level in string theory, including all the $\alpha'$ effects which distinguish string theory from pure supergravity in the bulk.  These are encoded in the $N_5$-dependence of the correlation functions.  From the expression for the string coupling, we see that there is a good perturbative string expansion even if $N_5$ is not large, by virtue of the large density.

Let us focus on the infrared region, treating the four $T^4$ directions as part of the space on which the dual field theory lives.  We may take a decompactification limit where the $T^4$ becomes simply $R^4$.   We may also consider semi-holographic couplings of the IR theory to other sectors of fields \cite{Semihol}, retaining significant control using the t'Hooft expansion of the holographic sector.

%Rather than matching to the UV little string theory, we can simply couple the system to a %dynamical potential field $B_{\mu\nu}$ semi-holographically.  This gives us a finite-%density system (of extended instanton strings) coupled semi-holographically to a dynamical% gauge field.

\subsection{Operators}

In the supergravity (SUGRA) limit, gravity-side fields are dual to CFT operators.  We will be concerned with string-theoretic effects that go beyond the SUGRA limit.
One can analyze perturbative string theory using a first quantized path integral, written schematically in terms of the spacetime metric $G_{MN}$ and the worldsheet metric $\gamma_{\alpha\beta}$ as
\beq\label{wsac}
\int [DX][D\gamma]e^{\frac{i}{\alpha'}\int d^2z \sqrt{-\gamma} G_{MN}(X)\gamma^{\alpha\beta}\partial_\alpha X^M\partial_\beta X^M +\dots} \,,
\eeq
where the worldsheet fields correspond to embedding coordinates $X^M$.  In a flat ``target" spacetime,  the metric $G_{MN}=\eta_{MN}$ is independent of the embedding coordinates, and this action is Gaussian in the $X^M$, leading to a relatively straightforward analysis.  But on curved spacetime backgrounds, $G_{MN}$ depends on the $X^M$ and this path integral is not Gaussian; there are nontrivial $\alpha'$ corrections.  For the $AdS_3$ case, the symmetries of the problem help, and a complete solution has been proposed \cite{Hobbit}\cite{Silmarillion}\cite{AdSthree}.  We will adopt these results to read off interesting properties of current correlators in the corresponding finite density field theory.

In the standard first-quantized description of perturbative string theory, single-string states are in one-to-one correspondence with integrated ``vertex operators" living on the 1+1 dimensional worldsheet of the string, the theory (\ref{wsac}) of two-dimensional gravity coupled to matter fields describing the embedding of the string in the ambient spacetime.
% [restated below]  These operators satisfy the constraint equations of the two-dimensional gravity, one of which requires that the vertex operator be dimension 2, half of which comes from left-movers and the other half from right-movers on the string.

There are many limits of string theory in which a perturbative description applies.  The simplest technically is the bosonic string theory in flat space or in a sufficiently symmetric background to be solvable, such as $AdS_3$.  This theory has an instability at the quantum level (a negative mass squared mode in the spectrum), but can be consistently analyzed at the classical level.  This case was worked out in detail in \cite{AdSthree}, the the superstring generalization is discussed further in \cite{Hobbit}\cite{AdSthreesusy}.  The essential physics of interest can be understood in either case: at the semiclassical level one finds singularities in correlation functions at finite $T^4$ momentum which are accounted for by string configurations that stretch out toward the boundary at zero cost in energy; this effect is common to the bosonic and fermionic strings.  It is interesting to note that the ``$2k_F$" singularities we are aiming for are related to fermion physics in condensed matter theory, but that we find the same kind of singularities from soft modes at nonzero momentum  also in the classical bosonic string.  But it is only the fermionic case which is stable, so it is only for that case that we have a well defined field theory dual.

We will write the results for the superstring following Ref.~\cite{AdSthreesusy}.  In comparing this to the bosonic results of \cite{AdSthree}, the two worldsheet models share a sector which is a bosonic $SL(2,R)$ WZW model at level $N_5+2$.  (In the notation of Ooguri and Maldacena \cite{AdSthree}, our $N_5$ translates to their $k-2$, with $k$ the level of the $SL(2,R)$ WZW model.)  Readers who wish to skip the technical background (which we will only be able to briefly summarize here in any case) can go directly to the results for current-current correlators discussed in section \ref{subsec:twopoint}\ below.

%{\bf ****In this section I need to finish adjusting the formulas for the superstring, which seems from the Rastelli paper to entail a small shift in $N_5$ as just indicated.  The notation should also be changed to not use $k$ for this.****}

The worldsheet CFT has a product target space $AdS_3\times T^4\times S^3$, so the vertex operators are schematically of the form\footnote{Here we describe the vertex operators in the ``zero ghost picture" for simplicity.}
\beq\label{vertex}
\int d^2z\, {\cal V}_{AdS} {\cal V}_{T^4} {\cal V}_{S^3}.
\eeq
These satisfy the worldsheet physical state conditions, generalizing the supergravity equation of motion.  Two of these conditions are that the
% zero-mode of the worldsheet stress energy tensor vanish for both left and right movers, which implies that the
total dimension of the vertex operator adds up to 1 for both left and right moving degrees of freedom on the string, i.e. that it form a ``(1,1) operator" on the string worldsheet.

The spectrum and correlation functions of this theory were first analyzed in full detail in the bosonic case in \cite{AdSthree},
and we will draw heavily from their results.  We consider first scalar operators on $AdS_3$ which are left-right symmetric in both the worldsheet conformal field theory and in the dual $CFT_2$: the various conserved currents can be constructed in terms of these. The vertex operators ${\cal V}_{AdS}$ that we will be interested in are indexed by two numbers, a continuous variable $j\in (\frac{1}{2},\frac{N_5+1}{2})$ and an integer $w$ \cite{AdSthree}. We will start by considering $w=0$, which will apply for a finite range of $R^4$ momentum $\vec q$: $0\le |\vec q|<q_*$; we will give the value of $q_*$ below.
In this range, the vertex operators for scalar operators have an $AdS$ component denoted $\Phi_j$ \cite{Silmarillion}\cite{Hobbit}\cite{AdSthree}, which is a scalar primary operator of the $SL(2,R)$ current algebra.

%with dual $CFT_2$ dimension $(j,j)$ and worldsheet dimension $(-j(j-1)/N_5,-j(j-1)/N_5)%$.

The dimension of the resulting operator in the dual $CFT_2$ is
\beq\label{dimreln}
\Delta=\Delta_L+\Delta_R=2j.
\eeq
The {\it worldsheet} scaling dimension $\Delta_{\rm ws}$ of the $AdS_3$ part of the vertex operator dimension is given by
\beq\label{wsdimreln}
\frac{\Delta_{{\rm ws}}}{2}=-\frac{j(j-1)}{N_5}
\eeq
where $N_5$ is the number of NS5-branes in the original brane construction.

The physical state condition requires that the $AdS_3$ contribution to the worldsheet dimension of the vertex operator be negative for positive values of the dual CFT dimension (\ref{dimreln}).  This is similar to flat spacetime string theory, where the part of the theory involving the temporal embedding coordinate $X^0$ is a nonunitary CFT which contributes negative-dimension operators of the form $e^{i\omega X^0}$ to physical vertex operators.  When combined with spatial momentum $e^{ik_m X^m}$ and other contributions to the vertex operators describing string oscillations, the physical state condition becomes the mass shell condition $\omega^2=\vec k^2+ {m}^2$.   The $AdS_3$ case is a generalization of this, for which the physical state conditions require
\beq\label{Lzero}
-\frac{j(j-1)}{N_5} + \frac{\Delta_{{\rm ws}, T^4}}{2}+\frac{\Delta_{{\rm ws}, S^3}}{2} = 1
\eeq
The first, negative, term here is the worldsheet dimension of the $AdS_3$ part of the (unintegrated) vertex operator (\ref{wsdimreln}).

An important contribution to $\Delta_{{\rm ws},T^4}$ comes from momentum $\vec q$ in the $T^4$ directions, in general:
\beq\label{qdim}
\Delta_{{\rm ws}, T^4}\sim \frac{\alpha'}{2} { q}^{2} + {\rm oscillator}
\eeq
where ``oscillator" refers to oscillator excitations in the $T^4$ directions.

As mentioned above, $j\in (\frac{1}{2},\frac{N_5+1}{2})$; one obtains dimensions outside the corresponding range of $q$ in (\ref{Lzero},\ref{qdim}) from an additional quantum number $w$ associated with spectral flow \cite{AdSthree}.  We will briefly describe those sectors below after first completing our discussion of the physics of the $w=0$ sector.

\subsection{Spacetime currents}

We would like to compute quantities such as the shear viscosity and current-current two point functions which are natural to consider in finite density field theory.  In this subsection we will explore the structure of currents in the $AdS_3\times T^4\times S^3$ string theory, focusing on the transverse components which will largely suffice for our application.  A systematic description of physical states and current components in our 5+1 dimensional finite density theory appears in the appendix; see e.g. \cite{Hobbit}\ for an analysis of currents in the 1+1 dimensional CFT dual.    

Consider a current $J^M$ in the full 5+1 dimensional field theory.  Its conservation requires
\beq\label{conservation}
\partial_{X_r}J^r+\partial_\mu J^\mu   = 0
\eeq
where $r$ labels the $T^4$ directions and $\mu$ the 1+1 spacetime directions occupied by the $CFT_2$ dual to $AdS_3$.  At zero momentum $q^r$ along the $T^4$ directions, these currents reduce to currents in the $CFT_2$, satisfying $\partial_x \bar J=0=\partial_{\bar x} J$, explained in \cite{Hobbit}.  We would like to understand these operators at nonzero momentum $q^r$.

One example is the stress-energy tensor $T_{MN}$, which in the (super)gravity limit is dual to the metric perturbations $G_{MN}$.  The two-point function of its components $T_{rs}$ along the $T^4$ directions at zero momentum determines the shear viscosity in these directions \cite{KSS}.  For this operator, in the case its indices $rs$ are transverse to any momentum $\vec q$ on the $T^4$ directions, the internal worldsheet vertex operator is ${\cal V}_{T^4}\sim \partial X^r\bar\partial X^se^{i\vec q\cdot\vec x}+{\it fermions}$, with worldsheet dimension $\Delta_{{\rm ws},T^4}=2(1+\alpha' q^{2}/4)$.
%\footnote{This is in the 0 ghost picture.  Interestingly, many of the features we will %discuss happen also in the bosonic string theory, if one ignores its tachyon %instability.}
This leads to
\beq\label{jq}
\frac{j(j-1)}{N_5} = \frac{\alpha'  q^{2}}{4} \Rightarrow j=\frac{1}{2}\left(1+\sqrt{1+N_5{ q}^{2}\alpha'}\right),
\eeq
which agrees with the usual $AdS_3/CFT_2$ dictionary for a massive scalar in $AdS_3$, with $ q^{2}$ playing the role of the mass squared in the reduction to 2+1 dimensions (note that the $AdS$ curvature radius squared  $\sim N_5$ at large $N_5$). Recall (\ref{dimreln}) that $j$ is half the total (left+right) dimension of the operator in the dual $CFT_2$.  Similar comments apply to the string current $J_{MN}$ dual to the Neveu-Schwarz antisymmetric tensor field $B_{MN}$.  

The vertex operator for transverse $T^{rs}$ is straightforward:
\bea\label{Trs}
\int d^2 z\, {\cal V}^{rs}&=&\int d^2z \,\Phi_j \partial X^r\bar\partial X^s e^{i\vec q\cdot \vec X}  \,,\eea
where $\vec q,\vec X$ are along the $T^4$ directions and $\Phi_j$ is the scalar primary operator of the $SL(2,R)$ current algebra mentioned above, with (Left, Right) dual $CFT_2$ dimension $(j,j)$ and worldsheet dimension $(-j(j-1)/N_5,-j(j-1)/N_5)$.

In the un-spectral-flowed window of $j$ and $\vec q$ values, we have an explicit expression for $\Phi_j$ at the classical level on the worldsheet \cite{AdSthree}\
\beq\label{Phij}
\Phi_j=\frac{1-2j}{\pi}\frac{1}{(e^{-\phi}+|\gamma-x|^2e^\phi)^{2j}}
\eeq
%and one can differentiate this with respect to $z$ and then integrate with respect to $x$ %to satisfy (\ref{VAdS}).
in terms of the Poincare embedding coordinates
\beq\label{Poincmetric}
ds^2_{AdS_3}=d\phi^2+e^{2\phi}d\gamma d\bar\gamma \,.
\eeq
This is a string-theoretic generalization of the bulk-boundary propagator, which asymptotes to a delta function $e^{2(j-1)\phi}\delta^{(2)}(x-\gamma)$ at the boundary (up to the indicated power of the warp factor $e^{2(j-1)}$ determined by the scaling dimension $j$ of the corresponding $CFT_2$ operator ${\cal O}_j$).  In the supergravity limit, this normalization ensures that the bulk field reduces to the source $\phi_0$ for ${\cal O}_j$.  This remains true after including $\alpha'$ effects: the equations of motion get corrected in the bulk, but the boundary condition remains the same \cite{Silmarillion}.

\subsection{Scales}

The $CFT_2$ dual to $AdS_3$ is of course scale invariant.  Nonetheless,
in the 5+1 dimensional description of the theory, there is a dimensionful quantity (the density) in addition to dimensionless couplings that may be varied, the latter corresponding to moduli on the gravity side.  The dimensionful parameter comes about as follows, similarly to the $AdS_2\times R^2$ case \cite{MIT}\cite{AdStwo}.  Physical quantities are naturally expressed as functions of momentum $\vec q$.  In (\ref{jq}) -- and as we will see in many quantities that depend on it -- there appears the combination $ q^{2}/\tilde\mu^2$ where $\tilde\mu^2=1/N_5\alpha'$.  Using the relations (\ref{NSFonescalings}), we can relate $\tilde \mu$ to the density $\hat\rho_1$:
\beq\label{mutilde}
\tilde\mu^2\sim \frac{1}{\hat R^2} \sim \frac{g_s}{N_5^{3/2}}\sqrt{\hat\rho_1}.
\eeq
We will find singularities in correlation functions at various nonzero momenta, the first appearing at a value $q_*$ determined below; $q_*^2$ will also be proportional to $\sqrt{\hat\rho_1}$ but with a different dependence on $N_5$.

This is analogous to the appearance of the Fermi momentum scale $k_F$ in the low-energy theory of weakly interacting fermions at finite density.  There, if one considers small frequencies, this limits one to the region near the Fermi surface.  In this infrared regime, locally on the Fermi surface the physics does not retain any dependence on $k_F$. However one can detect the scale $k_F$ via low energy probes at large momentum, in particular from the $2k_F$ singularities in correlators such as the density-density two point function.  With small interactions, one finds additional singularities at higher multiples of $k_F$.

Before continuing, let us note that
this is also analogous to the $AdS_2\times R^2$ theory considered in \cite{MIT}\cite{AdStwo}.  There, if one UV completes the theory with a 2+1 dimensional CFT dual to $AdS_4$, one can relate the analogue of $\tilde\mu$ arising in that theory to the chemical potential of the UV CFT.  However, one may also work directly with the theory dual to $AdS_2\times R^2$, perhaps coupling it semi-holographically to other fields \cite{Semihol}.  In that case, one again detects the scale $\tilde \mu$ by its appearance in the correlation functions written in terms of the $R^2$ momentum.

%With momentum $\vec q$ in the $T^4$ directions, this corresponds to a massive 2-form potential field in $AdS_3$ of mass $|\vec q|$.  For a $p$-form in $AdS_{d+1}$, the relation between mass squared $\vec q^2$ and dimension $\Delta$ is
%\beq\label{pformdim}
%\Delta = \frac{1}{2}(d\pm \sqrt{(d-2p)^2+4\vec q^2R_{AdS}^2}).
%\eeq
%(REF MAGOO review) in terms of the $AdS_{d+1}$ curvature radius $R_{AdS}$.  For our case of $d=2$ and $p=2$, this reduces precisely to the formula (\ref{jq}) for massive scalars.  This makes sense because a massive 2-form potential field in 2+1 dimensions has one scalar degree of freedom.  The gauge-invariant Stuckelberg mass term is of the form $(B_{\mu\nu}-\partial_\mu C_\nu)^2$.  Shifting $B_2\to B_2+d\Lambda_1$ is absorbed by $C_1\to C_1-\Lambda_1$.  This allows us to fix the gauge to $\partial^\mu B_{\mu\nu}=0$, so that the physical degrees of freedom have polarization $\epsilon_{\mu\nu}$ satisfying $q^\mu \epsilon_{\mu\nu}=0$, leaving one scalar degree of freedom.  In the massless case, i.e. with zero momentum along the $T^4$, the $B_{\mu\nu}$ field has zero degrees of freedom.  This is in accord with the fact that the instanton strings are spacefilling in the 1 spatial dimension dual to $AdS_3$; one cannot change the density locally in this direction
%without introducing $\vec x$ dependence along the $T^4$.

\subsection{Two-point functions, viscosity, and singularities}
\label{subsec:twopoint}

As explained extensively in \cite{AdSthree}, the correlation functions of our model can be computed explicitly.  For simplicity, we will focus on the components of the two point Green's functions which are related to transport and which have been studied in weakly coupled fermion systems to exhibit $2k_F$ singularities.

In momentum space, at zero temperature the normalized imaginary part of the retarded 2-point Green's function for scalar operators ${\cal O}$ with dimension $\Delta=2j=1\pm\sqrt{1+N_5{ q}^{2}\alpha'}$ takes the form
\beq\label{Twoq}
{\rm Im}(G^R_j)|_{T=0, scalar}=C(\omega^2-q_\sigma^2)^{2j-1}\theta(\omega^2-q_\sigma^2)  \sin(2\pi j)\frac{\Gamma[2-2j]}{\Gamma[2j+1]} (2j-1)^2\times \hat B(j) \,.
\eeq
The coefficient $C$ can be related to the basic parameters of the model; we will do this shortly.

As discussed above, we are particularly interested in the two point functions of currents in our dual 5+1 dimensional field theory at finite density.\footnote{This is partly for convention's sake; singularities associated with soft modes at nonzero momentum appear in the two point functions of general operators.  So any transport measurement would manifest such modes.}  The shear viscosity is given by the two point function of $T_{rs}$ (with $r\ne s$ both along the $T^4$ directions), at zero momentum.  This is a scalar operator from the point of view of the 1+1 CFT dual to $AdS_3$, so it behaves as in (\ref{Twoq}).
The transverse currents $J^{rs}$, with $q^r=0=q^s$ also correspond to scalar operators with the two point function (\ref{Twoq}).  

Now let us discuss the coefficient $C$.  It is proportional to $1/g_s^2$;
%\beq\label{Ct}
%C_t\sim \frac{1}{g_s^2}\tilde\mu^4
%\eeq
%where $\tilde\mu\sim 1/\hat R$ (\ref{mutilde}).
this scaling of this coefficient with the string coupling $g_s$ follows from the $({\cal R}+|H_{(3)}|^2)/g_s^2$ term in the gravity-side Lagrangian (\ref{action}).  This is similar to the factor $C\propto L_{AdS_4}^2/\kappa_4^2$ in the Reissner-Nordstrom black brane case \cite{AdStwo}, which scales like the number of field theoretic degrees of freedom of the field theory dual (i.e.\ the area in Planck units of an AdS radius size region \cite{Susskind:1998dq}).  We will comment on this further below in comparing our results to Luttinger's theorem.

These two point functions include the factor
\beq\label{norm}
\hat B(j) = \left(\frac{\Gamma[1-\frac{1}{N_5}]}{\Gamma[1+\frac{1}{N_5}]}\right)^{1-2j}
\frac{\Gamma[1-\frac{2j-1}{N_5} ]}{\Gamma[1+\frac{2j-1}{N_5}]}
\eeq
The $N_5$ dependence of the two point function encodes stringy effects, and through the ratio $(2j-1)/N_5$ with $j$ given by (\ref{jq}) it encodes nontrivial momentum dependence on the $T^4$ which is also stringy in origin.  We will discuss these effects further below.  
%Even at the level of supergravity (where we ignore $1/N_5$ corrections), the momentum dependence on the $R^4$ does %not admit a Fourier transform; the operators are not local.
%This is similar to the much-studied $AdS_2\times R^2$ theory, where the same comment applies to the momentum along the %$R^2$ directions.
%\footnote{In particular, it has nothing to do with the fact that the NS5-brane ``little string theory" is not a local field %theory.})
%Therefore we will continue to work in momentum space along these directions. 
 In the supergravity limit, the formula (\ref{Twoq}) reduces to the corresponding normalized two point function \cite{Normalized}.  As mentioned above, the normalization was fixed more generally by Teschner \cite{Silmarillion}.

%The factor $\hat B(j)$ is a nontrivial dynamical result, not determined purely by conformal symmetry.  It is related to the %reflection amplitude for stretched strings to bounce off the boundary.
%This provides a string-theoretic generalization of the standard supergravity dictionary in AdS/CFT relating the two-point %function to the relative coefficient of the two leading terms in the expansion of a bulk field near the boundary.

The corresponding finite temperature result for the scalar operators (\ref{Twoq}) is then given by
\bea\label{TwoqT}
{\rm Im}(G^R_j)|_{scalar} &=& \frac{2 C}{(2\pi)^2}\times e^{4j-2} (4\pi T)^{4j-2} \nonumber\\
& \times &\frac{\Gamma[2-2j]}{\Gamma[2j+1]}(2j-1)^2\sin(2\pi j)\theta(\omega^2-q_\sigma^2)~ {\rm sgn}\,\omega \nonumber\\
&\times & \hat
B(j)\sinh(\omega/2T) \bigl|\Gamma[j+ip_+/2\pi T] \Gamma[j+ip_-/2\pi T]\bigr|^2 \,.
\eea
Here $p_\pm=(\omega\pm q_\sigma)/2$, with $\sigma$ the spatial direction along the boundary of $AdS_3$.
In general, the factor $\hat B$ sits in front of all the two point Green's functions.

%In position space in time and $\sigma$, the finite-temperature 2-point function of scalar operators ${\cal O}$ on $AdS_3$ with dual-CFT dimension $\Delta=2h=2j$ takes the form
%\beq\label{Two}
%\langle {\cal O}(x){\cal O}(0)\rangle =  \tilde B(j) \frac{(\pi T)^{4j}}{sinh^{2j}(\pi T %x_++\pm i\epsilon)sinh^{2j}(\pi T x_-\pm i\epsilon)}
%\eeq
%where $x_\pm=\tau\pm \sigma$ are the dual CFT coordinates and the epsilon prescription %determines the precise Lorentz-signature Green's function to be computed.  This reduces to $\tilde B(j) \frac{1}{(x_+x_-)^{2h}}$ at zero temperature.

%Here the normalization includes the factor
%\beq\label{norm}
%\tilde B(j) = \frac{2j-1}{\pi}\left(\pi\frac{\Gamma[1-\frac{1}{k-2}]}{\Gamma[1+\frac{1}{k-2}]}\right)^%{1-2j}
%\frac{\Gamma[1-\frac{2j-1}{k-2} ]}{\Gamma[1+\frac{2j-1}{k-2}]}
%\eeq

The normalized 2-point functions exhibit very interesting features at zero momentum and at a nonzero momentum $q_*$:

\subsubsection{$\vec q \to 0$ and viscosity} 

From (\ref{jq}), $j=1$ at $\vec q=0$.  Plugging this into (\ref{norm}), we see that the $N_5$ dependence is completely eliminated; the $\Gamma$ functions with $N_5$-dependent arguments in (\ref{norm}) completely cancel out.  Thus, for zero-momentum transport calculations, including the viscosity to entropy ratio $\eta/s$, the supergravity result is exact.  

The absence of $\alpha'$ corrections\footnote{Note that $\alpha' /\hat R^2\sim 1/N_5$, Eq.~ (\ref{NSFonescalings})) .} was found previously for the NS5 system but at vanishing F1 density from the spectrum of hydrodynamic modes~\cite{Andreis}.  We can carry out the same analysis here.\footnote{We thank Andrei Parnachev for suggesting that we look at the hydrodynamic modes.  See \cite{viscosity}\ for a review of the corrections to $\eta/s$ in other examples.}
Poles at $\omega \to 0$ can arise only from the final $\Gamma$ function factors in the finite temperature two point function (\ref{TwoqT}).  These are at
\beq\label{soundpole}
(\omega \pm q_\sigma) = -4\pi i T h
\eeq
where $h$ is the spacetime (dual $CFT_2$) weight (for left movers).  So to get a hydrodynamic mode we need an operator whose $h$ (or $\tilde h$) vanishes when $q$ goes to zero. 
In the Appendix, we construct the states corresponding to more general components of the currents, going beyond the simplest, transverse components we have so far used.
Among these, the components $T_{\bar x\bar x}$ and the transverse part of $T_{\bar x r}$ have the property we need, left-moving dimension $j-1$.   For these,
\beq
j(j-1)/N_5 = \alpha' q^2/4   \to j-1 \sim \alpha' N_5  q^{2}/4.
\eeq
At $q_\sigma = 0$ this gives the dispersion relation
\beq
- i \omega = \pi \alpha' T N_5  q^{2} \,.  \label{hydro}
\eeq
For $q_\sigma$ nonzero, the behavior is not hydrodynamic, due to the infinite set of conserved quantities in the CFT.

As explained in \cite{Andreis}, this is to be compared with the hydrodynamic relation
\beq
- i \omega (\epsilon + P) = \eta  q^{2} \,.  \label{hydro}
\eeq
The rest energy density of the F-strings, $\epsilon = N_1 / 2\pi\alpha' V_{T^4}$, dominates the pressure, and so
\beq
\eta = T N_1 N_5 / 2V_{T^4} \,.
\eeq
The central charge is $6 N_1 N_5$, giving an entropy density $S/V_1  = 2\pi T N_1 N_5$, and so 
\beq
\eta V_{T^4}  V_{1} /S = 1/4\pi
\eeq
This agrees with the result from the zero-temperature two point function; the supergravity result is uncorrected.  To be precise, this holds for the velocity of the fluid, and the shear $\sim\vec q$, both along the $T^4$.  This corresponds to the correlator of $T_{rs}$ considered above, and to the hydrodynamic pole in the correlator of the transvere part of $T_{\bar x r}$.

When the motion of the fluid is parallel to the F-strings (and the shear still on the $T^4$), the relevant $\epsilon$ and $P$ are only from the CFT excitations, and so of order $T^2$,
\begin{equation}
\epsilon = P = \pi T^2 N_1 N_5/ V_{T^4} \,.
\end{equation}
By symmetry, $T_{01}  = T_{xx} - T_{\bar x\bar x}$ should couple to these modes, and we observe the same hydrodynamic behavior~(\ref{hydro}) as before.  This then gives 
\begin{equation}
\eta = 2\pi^2 \alpha' T^3 N_1 N_5^2/ V_{T^4}\, ,\quad \eta V_{T^4}  V_{1} /S = \pi \alpha' T^2 N_5 \,.
\end{equation}
That is, the viscosity to entropy ratio goes to zero with $T^2$.  We can verify this surprising conclusion from the Kubo formula.  Current conservation relates $T_{xx}$ to the longitudinal part of $T_{xr}$, whose dimension at $\vec q = 0$ is $(2,1)$.  For this current the factor $T^{4j - 2}$ from the correlator (\ref{TwoqT}) becomes $T^{2(m + \tilde m - 1)} = T^4$.  Combined with the $T^{-1}$ from $\sinh \omega/2T$ this gives a viscosity of order $T^3$, as found from the hydrodynamic mode.

Of course, this is a rather exotic and anisotropic system: we might say that these are the most perfect liquid crystals.  The anisotropy seems to play an essential role here.  It is as though the flows carried by the different F-strings do not interact strongly, so there is no large resistance when the velocity has a gradient perpendicular to the strings.

The shear viscosity tensor was previously studied in anisotropic holographic systems beginning in Refs.~\cite{Landsteiner:2007bd}\cite{Natsuume:2010ky}.  In some systems \cite{Erdmenger:2010xm}\cite{Basu:2011tt}\cite{Erdmenger:2011tj}\cite{Rebhan:2011vd} nonuniversal values for certain components of this tensor have been found.  In particular, Ref.~\cite{Rebhan:2011vd} found result very similar to ours in an anisotropic plasma based on dissoved branes \cite{Mateos:2011ix}.   The mixed components of the viscosity tensor lie below the usual universal bound, and fall with decreasing temperature.

%{\bf ***This statement (and the above normalization) depends on Teschner's 9906215 appendix A (sec 5), where he claims %to fix the normalization completely.***}

\smallskip

\subsubsection{Large-$\vec q$ singularities}

There is an interesting divergence in the 2-point functions, coming from the factor $\Gamma[1-\frac{2j-1}{N_5} ]$ in (\ref{norm}).  This occurs when $j\to \frac{N_5+1}{2}$, which means that $ q^{2}$ approaches
\beq\label{qstar}
q_*^2= \frac{1}{\alpha'}\left(\frac{N_5}{4}-\frac{1}{4N_5}\right)\sim g_s\sqrt{\hat\rho_1/N_5}\left(\frac{N_5}{4}-\frac{1}{4N_5}\right) \,.
\eeq
This is the upper limit of the region of momenta we are considering so far; see the next subsection for a brief discussion of larger bands of momenta obtained via the spectral flow procedure derived in \cite{AdSthree}.
We can turn  (\ref{qstar}) around and describe the control parameter $g_s$ in field theory terms as the ratio $q_*^2/\sqrt{\hat\rho_1}$ times a function of $N_5$ (which we hold fixed).

This divergence has a simple explanation on the gravity side\ \cite{SeibergWitten}\cite{AdSthree}, where it can be seen to come from long circular strings which can stretch out to the boundary at a finite cost in energy of order the string tension times the AdS radius; the attraction to the center of $AdS$ is exactly balanced by the electric field pushing out on the strings.  The field theory side is very strongly coupled, and we do not know how to derive this effect independently using field theory techniques.

In Fermi liquids, $2k_F$ singularities in the density-density correlator arise from particle-hole pairs.  It is interesting to compare (\ref{qstar}) in our system to that, to see if (and in what sense) it may be analogous.  Let us first more generally consider at what level one might expect such singularities to be visible, i.e. whether they require string-theoretic and/or finite $N$ quantum physics on the gravity side.  The most obvious way to expose particle-hole pairs is at one loop on the gravity side, and that may be required in many examples.   In some examples -- such as ours -- an analogue of particle/hole pairs arises already in classical string theory, i.e. including all $\alpha'$ effects but not requiring loop effects on the gravity side.  Another such example may arise from DBI flavor branes \cite{KulaxiziParnachev}\cite{HPST}\cite{unpubstr}.  There, particle/hole pairs would be analogous to the positive and negative charges on the end points of open strings on the flavor D-brane.  Moreover, at finite density the D-brane electric field grows large, reducing the effective open string tension, making such strings energetically accessible even at relatively low scales.  It may even happen that in some set of examples, $2k_F$ singularities arise already at the classical supergravity level; some indications from entanglement entropy calculations suggest this according to recent works such as \cite{entanglement}.

In all cases, one may estimate the relevant scale at which momentum singularities will occur if a Luttinger relation holds, as we will next discuss.  One final general remark is that one may be able to isolate the physics leading to the singularities in correlators without necessarily computing the full correlation function.  (In the present F1/NS5 case, the singularity is understood in a much simpler way than the full solution to the model, though both have been determined.)  It would be interesting to pursue this for all stable examples, to check whether ``$2k_F$" singularities generally persist at strong coupling.

%Let us begin with the present example, and then we will draw some more general lessons.

In our F1-NS5 example, one obvious complication in making the comparison is that our density is a density of extended strings rather than particles.  Charge neutral excitations in our system which may be analogous to particle-hole pairs include the circular strings described above, not just string-antistring pairs.
%In the original NS5/F1 brane construction, one has a density of strings stretched along the spatial direction of what %becomes the Poincar\'e slicing of $AdS_3$.  A particle-hole pair would be analogous to a pair of oppositely oriented strings.  %The circular long strings that cause the divergence just mentioned in the two point function are net charge neutral in the %bulk; their winding number is not conserved since they can shrink, a feature of the dynamics discussed extensively in %\cite{AdSthree}.  They can be thought of as a particle-hole pair, with the oppositely oriented strings annihilating aside from %a loop in the middle where the two separate, forming a charge-neutral circular string.
In the next section, we will consider sectors involving multiply wound loops, which may perhaps be viewed as analogues of multiple particle-hole pairs.  Before that, we will compare to Luttinger's theorem in order to explore further the analogy between our singularity at $q_*$ (\ref{qstar}) and $2k_F$ singularities.

The Luttinger theorem and its generalizations play a central role in condensed matter theory \cite{Luttinger}.  Let us describe the relation and how it may be used to predict the scale of ``$2k_F$" singularities if they persist to strong coupling.\footnote{For other recent discussions of the Luttinger theorem and holography, see for example \cite{holoLuttinger}.}  For pointlike quasiparticles (as opposed to our density of extended strings), this relates the volumes $V_f$ of the Fermi surfaces of a model's elementary fermion fields to the charge density $\langle Q\rangle=$(average number of charges)/volume under a global symmetry. There is one relation for each global $U(1)$ symmetry \cite{Luttinger}.
\beq\label{Luttinger}
\langle Q^a\rangle =\sum_f Q^a_f V_f 
,,
\eeq
where $Q^a_f$ is the (integer) charge of the $f$th sector of fermions with respect to the $a$th global $U(1)$ symmetry.  In a holographic model, we may expect the number of fermion species $\sum_f 1$ to be of order the $AdS$ area in Planck units, which is the number $N_{\rm dof}$ of field theoretic degrees of freedom.  If the global charge of each is also of order 1, this gives a prediction $k_F^{d-1}\sim \langle Q^a\rangle/N_{\rm dof}$.  For Einstein/Maxwell models such as \cite{MIT}\cite{AdStwo}, this leads to a supergravity scale for $k_F$, of order $\sqrt{g_{00}}/R_{AdS}$, in terms of the gravitational redshift factor $g_{00}$ at the boundary between the UV theory and the IR quantum critical point based on the $AdS_2\times R^2$ geometry.  However, the effect need not be seen in classical supergravity -- in fact it has not been seen at that level \cite{AdStwo}.  To expose particle/hole pairs may intead require loops in the bulk.  It may be possible to exhibit the singular behavior of this loop diagram, something to which we hope to return in future work.

In our case, the theory is strongly coupled and the density consists of extended strings, so it is far from a Fermi liquid built from pointlike charges.  However, it is interesting to compare (\ref{qstar}) to the relation (\ref{Luttinger}). F1 strings are electric sources for the Neveu-Schwarz $B_{MN}$ field on the gravity side; this corresponds to a global symmetry in the dual theory.  For a single fermi surface built from unit charged fermions, this has the form $<Q>\sim N_f V_f$ with $N_f=\sum_f 1$. In our system, the role played by the number $N_f$ of unit charged ($Q_f=1$) fermion species in (\ref{Luttinger})  is played by the continuous parameter $1/g_s^2$, that is (\ref{qstar})
\beq\label{FoneLutt}
V_{q_*}\equiv q_*^4\sim {g_s^2} \hat\rho_1.
\eeq
This identification of $1/g_s^2$ as an effective number of degrees of freedom appears to arise also in the following independent way.  For the particle case, we can read off the number of species from the normalization of the two point function; explicitly in the Reissner-Nordstrom models this scales like the number of field theoretic degrees of freedom $N_{\rm dof}\sim L_{AdS}^2/G_N$ (times a function of the chemical potential, frequency, and momentum) \cite{AdStwo}.  Now
recall that as discussed below (\ref{Twoq}), in our case the density two-point function scales like $1/g_s^2$ (times a function of $\tilde\mu,\omega, q_\sigma$, and ${\vec q}$).   Given $N_f\sim 1/g_s^2$, it seems that
(\ref{qstar}) does give an analogue of the Luttinger relation in our system.  However, we do not understand this well from first principles.  It would be interesting to flesh this out by connecting it to physics in a weakly coupled limit of the theory, such as the symmetric orbifold point of 1+1 dimensional sigma model dual to $AdS_3$.\footnote{The $N_5$ dependence here is also interesting; it is tempting to incorporate it into the density factor, generalizing $\hat\rho_1\sim N_1/L^4$ to $\hat\rho_{15}\sim N_1N_5/L^4$.  This may be motivated from the fact that at the simplest, symmetric orbifold point, the sigma model target space is $(T^4)^{N_1N_5}/S_{N_1N_5}$.}

In comparing to the $2k_F$ singularities arising in weakly coupled fermionic systems, it is also interesting to ask about the effect of finite temperature $T\ne 0$ and the effect of a UV cutoff.  At weak coupling, the Fermi surface is smoothed out at $T\ne 0$.  In our system, at finite temperature there is an additional force attracting the strings to the center of $AdS$, which goes in the right direction to suppress the long string states just mentioned. Still, without an ultraviolet cutoff on our theory, the singularities remain at nonzero $T$ (as can be read off from (\ref{TwoqT})).  This happens because the effects of the temperature go away sufficiently close to the boundary, and the long string divergence remains.  However, with a finite UV cutoff (which is required in any case if one wishes to mock up a condensed matter system), the boundary is removed or at least drastically changed.  So with a finite UV cutoff, we expect that our singularities are smoothed out at finite temperature.  Even at zero temperature, with a finite UV cutoff the divergence in the two point functions will go away, turning into a peak which is perhaps similar to those of the Bose Metal studied in \cite{Luttinger}.  One can also cut off the divergence from the long strings by introducing various kinds of axions on the $T^4$  \cite{SeibergWitten}.

%The momentum also comes into the two point function in the power of $\omega^2-q_\sigma^2$ in (\ref{Twoq}); at higher %momentum $q$, this power increases and the effect becomes weaker at small frequency.  This may also be interesting to %compare to the case of weakly interacting Fermions. (Pepin REF?)

%Recall that so far we are considering the range $1/2 < j<(N_5+1)/2$, with the other quantum number $w$ taken to be zero; the two-point function grows without bound as we approach the upper boundary of this region.  But we cannot simply analytically continue past the pole according to the Maldacena/Ooguri analysis; we must use their spectral flowed representations which build up a unitary, ghost-free theory.)

\subsubsection{Additional structure at larger momentum}

So far, we have seen an interesting divergence in two point functions as $|\vec q|\to q_*$, (\ref{qstar}).  As described in \cite{Hobbit}\cite{AdSthree}, this comes from ``long string" configurations in which the string stretches out to the boundary with finite cost in action; in the Lorentzian theory the short strings which are discrete bound states in $AdS_3$ become degenerate with long strings.

Above this value of momentum, there is an important subtlety in the computation of the spectrum and operators of the theory, the implications of which are laid out in the trilogy \cite{AdSthree}.  In order to obtain a unitary theory, one finds that the $AdS$ part of the vertex operators must now be built from a vacuum state which is annihilated by a shifted (``spectral flowed") subset of the modes of the currents in the worldsheet SL(2,R) WZW model.  Let us briefly explore the effect of this on the momentum dependence of correlators in bands of momenta with $|\vec q|>q_*$.

As mentioned above, we are interested in the lowest spacetime energy states which are worldsheet current algebra primaries.  We consider these lowest-lying states rather than excited oscillator levels because the latter have too many physical states to be dual to nontrivial conserved currents.  One subtlety is how to unambiguously continue the currents of interest into the higher bands of momenta, given the existence of two classes of operators (known as short/long strings, or discrete/continuous representations \cite{AdSthree}) both of which include oscillator ground states.  We will consider here the lowest energy states in each band of momenta, and read off from the solutions in \cite{AdSthree}\ where additional singularities in momenta can arise in the two point function of the corresponding operator.  This is sufficient to exhibit additional sectors of soft modes related to multiply wound strings (perhaps analogous to multiple particle-hole pairs).

Starting at $q_*^2$, there is a strip of width $1/2$ in the momentum squared $q^2\alpha'$ for which there are no discrete (short string) states at the lowest string oscillator level.  Instead, the lowest energy states are still long string states in this strip.  They are delta function normalized, and their two point functions have no poles, zeros or other singularities in this range.  For a finite range of momenta above that, a new window of discrete states contributes, starting at a zero at $q^2\alpha'\to (N_5+4)/4-1/4N_5$ and ending at a divergence as $q^2\alpha'\to N_5+1-1/4N_5$.

This structure repeats, with the successive ranges of momenta indexed by an integer $w$.  The strip with long strings as the lowest lying states grows larger (of width $w$, for increasing positive integers $w$), lying in the range
\beq\label{longrange}
\frac{(N_5+2)w^2}{4}-\frac{w}{2} < q^2\alpha' +\frac{1}{4N_5} < \frac{(N_5+2)w^2}{4}+\frac{w}{2} \,.
\eeq
The range of momenta for which the lowest lying operator is a short string is given by
\beq\label{shortrange}
\frac{(N_5+2)w^2}{4}+\frac{w}{2} < q^2\alpha' +\frac{1}{4N_5} < \frac{(N_5+2)(w+1)^2}{4}-\frac{w+1}{2} \,.
\eeq
%(This corresponds to $1/2 < j < (k-1)/2$ in the notation of Maldacena/Ooguri, where we committed to a state with no %oscillator excitations and with the worldsheet dimension from the internal $T^4$ directions being $h=1+q^2\alpha'$.)

The complete calculation of two point functions (as well as higher point functions) was developed in \cite{AdSthree}.  From that solution, we can read off for example the loci in momentum space where the two point function vanishes or diverges.  It approaches zero as we approach the following momenta from above
\beq\label{zero}
q^2\alpha'\to \frac{(N_5+2) w^2}{4}+\frac{w}{2}-\frac{1}{4N_5} ~~~~~~ w\ne 0  \,.
\eeq
This is the lower end of the $w$th band of momenta covered by the discrete (short string) states.
%\footnote{corresponding to $j\to 1/2$ in the notation of Maldacena/Ooguri.}

The two point functions diverge as we approach the following momenta from below (generalizing the behavior as $q^2\to q_*^2$ we discussed above):
\beq\label{infinity}
q^2\alpha' \to \frac{(N_5+2) (w+1)^2}{4}-\frac{w+1}{2}-\frac{1}{4N_5} \,.
\eeq
This is the upper end of the $w$th band of momenta covered by the short string states.%\footnote{corresponding to $j\to (k-1)/2$ in the notation of Maldacena/Ooguri.}

For weakly coupled fermions, there is also spectral weight above $2k_F$, with additional singularities arising at multiples of $2k_F$ from multiple particle/hole pairs.  We are at strong coupling, and it would be interesting to understand the pattern of singularities we get here from a more traditional field theory point of view.

The two point functions in this theory seem not to be analytic in the momentum along the $R^4$ directions; indeed the spectral flow discovered in \cite{AdSthree}\ avoids nonunitary behavior which would follow from a naive analytic continuation of the two point function (\ref{Twoq}).  Usually one derives the analytic structure of two point functions from a spectral decomposition, introducing a complete set of states which includes an integral over all momenta.  In our case, the complete set of states one inserts is itself described in the piecewise way just indicated, depending on the integer $w$ whose value depends on which range of momentum one considers.  It would be interesting to understand better how this arises from the point of view of the 5+1 dimensional dual theory.

%Are there any theorems about the required analytic structure of correlation functions? I don't see how the Maldacena/Ooguri ones are just the analytic continuation of the above to arbitrary $j$; it seems that they are making a big point that their spectral flowed correlation functions are not obtained that way.

\subsubsection{Remarks}

There have been previous indications that stretched strings could produce ``$2k_F$" type singularities. One example is the DBI theory of flavor branes in gauge/gravity duals, for which a critical electric field develops which effectively reduces the open string tension, making stretched strings energetically cheap.  The present work seems encouraging for this idea more generally, and it would be interesting to return to other examples.  Relatedly, it would be interesting to understand whether there is a useful string bit picture which explains the Fermi surface like behavior here in terms of fermionic constituents of larger gauge-invariant operators, as in discussions of Bose metals and FL* phases \cite{Luttinger}.

\section{Discussion}
\setcounter{equation}{0}

The main observations in this paper have been the following:

\noindent 1)  Large density can drive a controlled holographic description of a quantum field theory, even in some cases without a large number of colors.

\noindent 2)  One example of this is the infrared limit of the NS5/F1 theory, which can be viewed as a 5+1 dimensional theory at finite density.  This is stable, and exactly solved at the level of $\alpha'$ effects in string theory on the gravity side \cite{AdSthree}.  This solution yields three interesting results for questions of interest in finite density field theory:

a)  For velocities perpendicular to the strings, the viscosity to entropy ratio is $1/4\pi$, uncorrected by $\alpha'$ effects in this model.

b)  For velocities along the strings, the  viscosity to entropy ratio vanishes as $T^2$ at low temperature.

c)  There are singularities from soft modes at a sequence of nonzero momentum scales.

\noindent In particular, this provides a relatively clean example in ``AdS/CMT" where string theoretic effects are important to include and have been calculated.

\subsection*{Acknowledgements}
We thank X. Dong, T. Faulkner, M.P.A. Fisher, S. Hartnoll, and A. Parnachev for very helpful discussions and explanations of some relevant background.  We are also grateful to S. Kachru, J. Maldacena, D. Mateos, J. McGreevy, A. Rebhan, E.  Shaghoulian, S. Shenker, and D. Trancanelli  for useful comments.  E.S thanks the KITP for hospitality during the Holographic Duality and Condensed Matter Physics program.  This research was supported in part by the National Science Foundation under grants  PHY07-56174, PHY07-57035 and PHY11-25915, and by the DOE under contract DE-AC03-76SF00515. 

\appendix
\section{Conserved currents}
\setcounter{equation}{0}

In this appendix, we construct physical states and relate them to currents in our 5+1 dimensional finite density theory.  See \cite{Hobbit}\ for an analysis of currents in the 1+1 spacetime CFT.  In this appendix, for simplicity we focus on the bosonic theory and use the notation of \cite{AdSthree}.  

\subsection{$w=0$}

Single-trace operators in the CFT are in one-to-one correspondence with single particle string states.  Here we identify the physical states of low dimension, and then single out those dual to conserved currents.  The states are representations of the left- and right-moving $\widehat{SL}_k(2,R)$ algebras
\begin{eqnarray}
\,[J^3_n,J^3_m] &=& -\frac{k}{2} n\delta_{n+m,0} \,, \nonumber\\
\,[J^3_n,J^\pm_m] &=& \pm J^\pm_{n+m} \,, \nonumber\\
\,[J^+_n,J^-_m] &=& -2 J^3_{n+m} + kn\delta_{n+m,0} \,,
\end{eqnarray}
and of the oscillator algebra on the $T^4$; we take all states to be trivial on the $S^3$.  The spacetime weights $m,\tilde m$ are the eigenvalues of $J_0^3, \tilde J_0^3$.  We consider massless states with $w=0$ (no spectral flow).  To obtain these, we start from a current algebra ground state $|j,m,\tilde m\rangle$ and excite to the first level on the right and left.  The mass-shell condition is then as noted in the text, $2 j(j-1) = (k-2) q^2$.  In this appendix we work in units of $\alpha' = 2$.

We solve the physical state conditions, for which we can focus on the left- and right-movers separately.  We can then condense the notation $|j,m,\tilde m\rangle \to |m\rangle$.  There is a unique state of weight $m = j-1$:
\begin{eqnarray}
 \| 0 \rangle = J^-_{-1} | j \rangle/(k-2)  \,,
\end{eqnarray}
with a normalization that will be convenient below.  This state  is physical: $(k-2) L_1\| 0 \rangle  = J^-_{0} | j \rangle + J^-_{-1} L_1 | j \rangle = 0$.
The notation $\| m \rangle$ will denote a physical state at the first excited level whose weight is $j-1+m$ and so goes to $m$ in the limit $q^2 \to 0$.

At weight $m=j$ there is
\begin{eqnarray}
\alpha J^-_{-1} |j+1\rangle + \beta J^3_{-1} |j\rangle + e_r \alpha_{-1}^r |j\rangle \,,
\end{eqnarray}
with physical state condition 
\begin{eqnarray}
\alpha C_j + \beta j + e_r q^r  = 0 \,.
\end{eqnarray}
We have defined
\begin{equation}
C_m^2 = (m+1)m - j (j-1)\,,\quad
J_0^+ |m\rangle = C_m |m+1\rangle \,,\quad 
J_0^- |m\rangle = C_{m-1} |m-1\rangle  
 \,.
\end{equation}
In particular, $C_j = \sqrt{2j}$.
By subtracting a null state $\propto L_{-1}|j\rangle$ we can set $\alpha = 0$, and so the physical states are $e_r \| 1\,r \rangle$ where
\begin{equation}
\| 1\,r \rangle = ( - q^r  J^3_{-1}  / j +  \alpha_{-1}^r ) |j\rangle\,.
\end{equation}

At weight $m=j+1$ the general state is 
\begin{eqnarray}
\alpha J^-_{-1} |j+2\rangle + \beta J^3_{-1} |j+1 \rangle +  \gamma J^+_{-1} |j\rangle + e_r \alpha_{-1}^r |j+1\rangle \,,
\end{eqnarray}
with physical state condition 
\begin{eqnarray}
\alpha C_{j+1} + \beta (j+1) + \gamma C_j + e_r q^r  = 0 \,.
\end{eqnarray}
Again we subtract a null state to set $\alpha = 0$, leaving as physical states
\begin{eqnarray}
\| 2 \rangle &=&  - C_j J^3_{-1} |j+1 \rangle/(j+1) +  J^+_{-1} |j\rangle \,,
\nonumber\\
\| 2\, r \rangle &=&  \left(- q^r   J^+_{-1} /(j+1) + \alpha_{-1}^r \right)|j+1\rangle \,.
\end{eqnarray}

The spacetime symmetry $J^+_0$ takes physical states to physical states:
\begin{eqnarray}
J^+_0 \| 0 \rangle &=&  - q_r  \| 1 \,r \rangle + \mbox{null} \,,
\nonumber\\
J^+_0 \| 1\, r \rangle &=& C_j  \| 2\, r \rangle + q^r \| 2\, \rangle /j \equiv -\| 2 \,r \rangle'   + \mbox{null}\,.  \label{j0+}
\end{eqnarray}
We have defined $\| 2\,r \rangle'=  -C_j  \| 2\,r \rangle - q^r \| 2 \rangle /j  $ 
so that, up to null states,
\begin{eqnarray}
J^+_0 \| 0 \rangle+  q_r  \| 1\,{r} \rangle &=& 0 \,, \label{j0+2}
\\
 J^+_0 \| 1\,{r} \rangle + \| 2\,{r} \rangle' &=& 0\,. \label{desc}
\end{eqnarray}

The generators $J^+_0, \tilde J^+_0$ act on local operators as $P_x, P_{\bar x}$.  The condition for a conserved current is therefore that the corresponding states satisfy
\begin{eqnarray}
J^+_0  | {\cal J}_{\bar x} \rangle + \tilde J^+_0  | {\cal J}_{ x} \rangle + q_r | {\cal J}_{ r} \rangle = 0 \,.
\end{eqnarray}
In fact, the relation~(\ref{j0+2}) allow us to construct chiral currents in which only one of  $J^+_0$ or $\tilde J^+_0$ appears.  At $q = 0$ this is the familiar result that all CFT currents are chiral, but it is surprising that this extends to nonzero $q$.  Thus we define the following conserved currents:
\begin{eqnarray}
|{\cal J}_{\bar x, x}\rangle &=& \| 0 ,\tilde 0 \rangle \,,\quad |{\cal J}_{r, x}\rangle = \| 1\,{r} ,\tilde 0 \rangle 
\nonumber\\
|{\cal J}_{\bar x, s}\rangle &=& \| 0 ,\tilde 1\,{s} \rangle \,,\quad |{\cal J}_{r,s}\rangle = \| 1\,{r} ,\tilde 1\,{s}\rangle 
\nonumber\\
|{\cal J}_{\bar x, \bar x}\rangle &=& \| 0 ,\tilde 2\rangle \,,\quad |{\cal J}_{ r,\bar x}\rangle = \| 1\,{r} ,\tilde 2\rangle 
\\[4pt]
|\tilde{\cal J}_{x,\bar x}\rangle &=& \| 0 ,\tilde 0 \rangle \,,\quad |\tilde{\cal J}_{r,\bar x}\rangle = \|0 ,\tilde  1\,{r}  \rangle
\nonumber\\
|\tilde{\cal J}_{ x,s}\rangle &=& \| 1\,{s} ,\tilde 0\rangle \,,\quad |\tilde{\cal J}_{r,s}\rangle = \| 1\,{s} ,\tilde  1\,{r}  \rangle 
\nonumber\\
|\tilde{\cal J}_{ x, x}\rangle &=& \| 2 ,\tilde 0\rangle \,,\quad \|\tilde{\cal J}_{r, x}\rangle = | 2 ,\tilde  1\,{r}  \rangle 
\end{eqnarray}
In ${\cal J}_{M,N}$, $M$ is the component and $N$ distinguishes the different currents.
The states  $\| 2\,r \rangle', \| \tilde 2\,{r} \rangle'$ do not give rise to distinct currents because they are descendants by~(\ref{desc}). 

 From the weights and symmetries we identify  (1)
$\tilde{\cal J}_{ x, x} \propto T_{xx}$, ${\cal J}_{\bar x,\bar x} \propto T_{\bar x \bar x}$;  (2) ${\cal J}_{\bar x,s} = {\cal J}_{s,\bar x}$ and $\tilde{\cal J}_{ x,s} = \tilde{\cal J}_{r, x}$ are the currents associated with (KK momentum $\pm$ winding number) on the $T^4$; (3) the symmetric and antisymmetric parts of ${\cal J}_{r,s} = \tilde {\cal J}_{s,r}$ are $T_{r,s}$ and the winding number flux on $T^4$; (4) ${\cal J}_{r,\bar x} \propto T_{r \bar x}$, $\tilde{\cal J}_{r, x} \propto T_{r x}$;  (5) ${\cal J}_{\bar x, x} = \tilde{\cal J}_{x,\bar x}$ should be the F1 density (note that its dimension goes to zero with $q$, reflecting the nonzero expectation value).  Thus, the three currents in each set arise from conservation of F1 density, conservation of $T^4$ momentum and winding, and conservation of CFT Poincar\'e energy-momentum.

The identifications just given appear to overcount, in that $T_{rx}$ is both the weight $(2,1)+ O(q^2)$ current ${\tilde{\cal J}}_{r, x}$ and also the weight $(1,0)+ O(q^2)$ KK momentum current.  The point is that the transverse and longitudinal parts of $T_{rx}$ along $T^4$ have different dimensions, and we must refine the identifications above.  This issue has already arisen in the $AdS_2 \times R^2$ analysis of Refs.~\cite{AdStwo}.  Only the longitudinal part of ${\tilde{\cal J}}_{r, x}$ appears in the conservation equation, and so we identify this with the longitudinal part of $T_{rx}$.  Similarly, $U(1)$ KK currents are constant along the KK directions, and so only the transverse part should be identified with that of $T_{rx}$.\footnote
{In Refs.~\cite{AdStwo}, the correlators contained sufficient powers of $q^2$ that the separation into longitudinal and transverse parts was nonsingular at $q=0$.   These powers arose due to the $0+1$ dimensional kinematics of their CFT factor.  We have not studied the corresponding issue here.}

%We are missing an identification for the conservation of the longitudinal part of ${\cal 
%J}^{r,\bar x}$; perhaps this is an emergent symmetry in the IR.

\subsection{$w \neq 0$}

Spectral flow does not take conserved currents into conserved currents, because the spacetime derivative $J^+_0$ flows to $J^+_w$.  Thus there is no natural continuation of the $w = 0$ currents to larger ranges of $q^2$.  We must extend the construction to nonzero $w$.  The construction above does not apply directly, because it assumes that the states $(J_0^+)^l|j\rangle$ are lowest weight in the current algebra, which is only true at $w=0$.  However, the extension seems quite simple.  Note that
\begin{equation}
\| 1\,{r}\rangle = - (q^r/q^2) J_0^+ \|0\rangle + (\delta^{rs} - q^r q^s /q^2) \| 1\,{r}\rangle + \mbox{null} \,.
\end{equation}
Now, choose {\it any} states $|\chi_0\rangle$, $|\chi_1^r\rangle$.  Defining
\begin{equation}
\|0\rangle = |\chi_0\rangle \,,\quad  \| 1\,{r}\rangle = - (q^r/q^2) J_0^+ |\chi_0\rangle + (\delta^{rs} - q^r q^s /q^2) | \chi_1^{r}\rangle  \,,  \label{triv}
\end{equation}
we have a conserved current 
\begin{equation}
J_0^+ \|0\rangle + q_r \| 1\,{r}\rangle = 0
\end{equation}
(tensoring with any right-moving state).  Generally one regards such automatically-conserved currents as trivial; what makes the usual Noether currents (including those above) nontrivial is that they are nonsingular at $q = 0$, unlike the gereric (\ref{triv}).  But the currents at $w \neq 0$ are defined only for $q^2$ greater than some minimum value, and there seems to be no distinction in this case between trivial and nontrivial currents.  Presumably the energy-momentum tensor in the underlying theory goes over to a sum over all such currents, dominated by that of lowest dimension.

\begingroup\raggedright\endgroup

%%%%%%%%%%%%%%%%%%%%%%%%%%%%%%%%%
\end{document}